\documentclass[a4paper,10pt,prl,twocolumn]{revtex4-1}

\usepackage{graphicx}% Include figure files
\usepackage{dcolumn}% Align table columns on decimal point
\usepackage{bm}% bold math

\def\prl{ Phys. Rev. Lett.}

\def\etal{{\it et al.\/}}

\newcommand{\ignore}[1]{}

\newcommand{\be}{\begin{equation}}
\newcommand{\ee}{\end{equation}}
\newcommand{\bea}{\begin{eqnarray}}
\newcommand{\eea}{\end{eqnarray}}

\newcommand{\ket}[1]{\left\vert #1    \right\rangle }

\begin{document}

\title{Superfast Cooling}

%\author{(alphabetically for now){\\}} \noaffiliation
\author{S. Machnes$^{1}$, M. B. Plenio$^{2,3}$, B. Reznik$^{1}$, A. M. Steane$^{3}$, A. Retzker$^{2}$}

\affiliation{$^{1}$  Department of Physics and Astronomy, Tel-Aviv
University, Tel Aviv 69978, Israel}
\affiliation{$^{2}$ Institut f{\"u}r Theoretische Physik, Universit{\"a}t
Ulm, D-89069 Ulm, Germany}
\affiliation{$^{3}$  QOLS, The Blackett
Laboratory, Imperial College London, Prince Consort Rd., SW7 2BW,
UK}
\affiliation{$^{4}$  Department of Physics, University of Oxford, Clarendon Laboratory, Parks Road, Oxford OX1 3PU, U.K.}
\date{\today}% It is always \today, today, but any date may be explicitly specified

\begin{abstract}
Currently laser cooling schemes are fundamentally based on the weak coupling regime. This requirement sets the trap frequency as an upper bound to the cooling rate. In this work we present a numerical study that shows the feasibility of cooling in the strong coupling regime which then allows cooling rates that are faster than the trap frequency with state of the art experimental parameters. The scheme we present can work for trapped atoms or ions as well as mechanical oscillators. It can also cool medium size ions chains close to the ground state.
\end{abstract}

% PACS, the Physics and Astronomy Classification Scheme.

\maketitle

{\em Introduction ---}
Laser cooling is the main tool that enables the exploration of low temperature phenomena in atomic physics, at nano and micro systems as well as the rapidly developing field of quantum technologies. Following the original ideas of Doppler cooling\cite{Hansch,Wineland1}, laser cooling has taken a central role in the physics of cold atoms and the number of ideas and their sophistication is growing continuously.
In the quest to propose cooling schemes that can cool to lower and lower temperatures at ever increasing rates the complexity and the efficiency of laser cooling has progressed a long way.

Two level systems(TLS) admit the Doppler cooling limit. This limit can be overcome by the Sisyphus mechanism either for free\cite{SisyphusExp,SisyphusThe1,SisyphusThe2} or trapped particles\cite{SisyphusBound}. The recoil limit can be broken by higher level systems by Velocity Selective Coherent Population(VSCP)\cite{Aspect} trapping or by Raman cooling\cite{Kasevich} for free particles and in their analog for bound particles, Dark State cooling\cite{MorigiDark}  and Raman side band cooling\cite{RamanSideBandCooling}. Cooling schemes for trapped particles can be applied for mechanical oscillators in the setup described in \cite{Ignacio2004,Rabl2009}.

However, all such cooling schemes rely on a weak coupling between the internal degrees of freedom (dof) and the ion's external motional state. This results in a clear separation of time scales between the two, allowing adiabatic elimination of the faster of the two dof-s. Unfortunately due to the weak coupling nature, the trapping frequency necessarily becomes the upper limit to the cooling rate and thus the achieved cooling rates are a few orders of magnitude lower than the trap frequency.

We here present a fundamentally novel way of cooling a system of trapped particles, achieving rates that are faster than the trapping frequency, and which grows with the laser intensity  with no fundamental upper limit. We demonstrate by means of numerics the proposed principle on a linear ion trap and show in detail how, using optimized sequences of coupling pulses, rapid cooling can be achieved. The proposed method is robust to fluctuations of the laser and may be adapted
to a very wide range of systems, this can be achieved, for neutral atoms \cite{Kasevich}, trapped ions at low temperature \cite{Cold Ion Trap}, and mechanical oscillators \cite{Ignacio2004,Rabl2009} for which the method can break the temperature limit imposed by the finite Q factor.

{\em The superfast cooling concept ---}
Consider a trapped system which is coupled to a TLS. The system is in an almost-harmonic potential, with engineered coupling between the external dof and the internal ones. For a standing wave configuration when the ion or atom is at the node we obtain in leading order the following Hamiltonian:
\begin{equation}
H/\hbar= \delta \sigma_z + \nu a^{\dagger}a+\eta\Omega\left(a^{\dagger}+a\right)\sigma_{\theta}\label{eq:H_simplified},
\end{equation}
where $\eta$ is the Lamb-Dicke parameter, $\nu$ is the trapping frequency, $\Omega$ is the effective Rabi coupling between the two internal levels, $a$ is the phononic annihilation operators and $\sigma_\theta$ is the pseudo spin operator in the direction $\theta$ in the $x,y$ plane. To cool the system, we would like to transfer energy from the motional to the internal dof., which can be dissipativly reinitialized via optical pumping. In other words, we wish to implement the red sideband-cooling operator, $a\sigma^{+}+a^{\dagger}\sigma^{-}$, \emph{\cite{Sideband-1,Sideband-2,Sideband-3}}.
In sideband cooling this term is generated from the $\hat{X}\sigma_{x}$ term after a rotating wave approximation (RWA), which requires a time which is longer than the trapping frequency, where $\hat{X}$ is the position operator.

A natural question to ask is whether we can cool, i.e. create the sideband-like term, in times which are faster than the trap frequency. A positive approach is by numerical optimization, but this requires an important input which is the initial `point', i.e., a guess of the cooling cycle. One may try to initiate the optimization with the schemes valid for the weak coupling regime such as sideband cooling or the dark state cooling schemes and extend them to strong couplings. This approach proves highly inefficient.
In our work we approach the problem from the opposite direction:
we choose as our initial guess a cooling scheme in the impulsive limit (infinitely fast), where we could reach the ground state infinitely fast if we neglect the higher order terms in the Hamiltonian. Starting from this point we optimize numerically to adapt finite couplings attainable in the lab.

We start by discussing the idealized case to gain insight into the optimization procedure.
Cooling at the strong-coupling limit is simple and extremely fast. In this limit there are various `points' from which we can start the numerical optimization.
We choose to start with the following `point'.
Consider the following argument which elucidates the underlying intuition of our work: The side band cooling term can be written as $a\sigma^{+}+a^{\dagger}\sigma^{-}=\sqrt{\frac{m\omega}{2\hbar}}\left(\hat{X}\sigma_{x}-\frac{1}{m\omega}\hat{P}\sigma_{y}\right)$. We already have a $\hat{X}\sigma_{x}$ coupling available to us (\ref{eq:H_simplified}). Thus if we can create a $\hat{P}\sigma_{y}$ term, we can generate the red sideband Hamiltonian using the Trotter decomposition \cite{Suzuki-Trotter}, neglecting some constants: $\left(e^{i \Omega \hat{X}\sigma_{x}dt}e^{i \Omega \hat{P}\sigma_{y}dt}\right)^{n}=
e^{i  \left(\hat{X}\sigma_{x}-\hat{P}\sigma_{y}\right)\Theta}$ when $dt\rightarrow0$ and $\Omega n dt=\Theta$; i.e., multiple short pulses of $\hat{X}\sigma_{x}$ and $\hat{P}\sigma_{y}$ will create the needed Hamiltonian. Therefore, if we had infinitely strong lasers and the $\hat{P}\sigma_{y}$ interaction, we would be able to cool instantaneously, at ${\Theta}=\pi$.

{\em $\hat{P}\sigma_{y}$ term ---}  Following the insight first described in \cite{Detecting in vacuum} we derive $\hat{P}\sigma_{y}$ as an effective Hamiltonian using $\hat{X}\sigma_{y}$ pulses which give the ion momentum, a period of free evolution in which the ion translates and finally a $-\hat{X}\sigma_{y}$ pulse, imparting equal and opposite momentum, stopping the translation. Furthermore, in the strong coupling regime the coupling is much larger than the phonon energy, $\eta\Omega\gg\nu$, allowing us to ignore the free evolution for the duration of the pulses. Setting $\delta=0$ yields a solvable Hamiltonian which enables us to get an exact result.
Formally, splitting (\ref{eq:H_simplified}) into $H_{free}=\hbar\nu \left(a^{\dagger}a+\frac{1}{2}\right)$
and $H_{pulse}=\hbar\Omega\eta\left(a^{\dagger}+a\right)\sigma_{y}$, setting $\delta$ to zero and making use of the Baker-Campbell-Hausdorff formula \cite{BCH-1,BCH-2,BCH-3}
we get a closed-form expression
\be
e^{\frac{-i}{\hbar}\left(t_{f}H_{free}+\hbar\eta t_{f}t_{p}\Omega\nu\hat{P}\sigma_{y}+\hbar\eta^{2}\nu\Omega^{2}t_{f}t_{p}^{2}\right)},
\label{effevtive}
\ee
which includes the desired $\hat{P}\sigma_{y}$ operator. We can now combine a $\hat{X}\sigma_{x}$ pulse with a $X\sigma_{y}$-wait-$X\sigma_{y}$ sequence implementing a $\hat{P}\sigma_{y}$ pulse and generate a red-sideband
operator. Note the we have a large phase term $\hbar\eta^{2}\nu\Omega^{2}t_{f}t_{p}^{2}$ that we will have to deal with. Initial numeric optimization assumed that idealized $P$ pulses are possible, with relatively small density-matrix cut-off sizes, which together allowed a much-accelerated computation procedure, and then further work adapted these sequences to the realities of achievable pulse strengths and validated the cooling processes with much larger d.m.-s. As expected, the idealized pulses resulted in even better performance than is presented in this letter.

{\em Numerical optimization of realistic scenarios ---}
Moving closer to a true model of an ion-trap, we need to move away
from the impulsive limit and re-introduce free evolution during pulsing,
as well as account for residual detuning and higher-order elements
in the Lamb-Dicke approximation. In this case the Hamiltonian is not solvable anymore and the BCH form employed for the $\hat{P}\sigma_{y}$ demi-pulse
does not apply. Thus a correction should be added to Eq. \ref{effevtive}. The free evolution during the pulse changes the commutation relation between the two pulses as the exponent in the second pulse is not the negative form of the exponent
in the first pulse - the free evolution does not change sign. This gives a crucial correction to Eq.\ref{effevtive}, which has to be taken into account. Here we have a major problem, as this correction cannot be taken into account in a purely numerical way since this will force us to introduce numerically the operation of the strong pulse which will require us to use a truncation at large phonon numbers.

Moreover, the $\hat{P}\sigma_{y}$ demi-pulse generation is deeply
connected to the $\left[\hat{X},\hat{P}\right]$ commutation relation,
which breaks once one tries to use finite-sized matrices in numerical
calculations, and the size cut-off introduces significant additional inaccuracies when
trying to make use of these commutation relations via $3$ matrix exponents.
We must therefore compute the equivalent to the pulse-wait-pulse sequence
analytically, merging the $3$ exponents into one, and only then can
we safely chain operations into a complete cooling sequence, which
is less sensitive to numerical inaccuracies. We therefore employ
the expanded BCH form as appearing in \cite{BCH for 3}.

Unlike the BCH form used to derive the effective $P$-pulse in the infinite limit, the BCH series does not terminate after a finite number of elements,
and has to be computed to high order to achieve accurate results.
To achieve a workable, reasonably-compact form for such a series requires
the use of computerized algebra which is capable of making use of commutation-relations
to simplify expressions. This has been implemented in Mathematica. This avoids errors due to $[x,p]$ numerical failure.

%While the effectiveness of the specific sequences discussed in the examples
%above is reduced when used with such expanded Hamiltonians, further
%optimization performed on these more realistic system models result
%in new sequences which exhibit very similar performance characteristics.

{\em Results  ---}
As the $\hat{P}\sigma_{y}$ interaction shifts the location of the ion it cannot be performed instantaneously. Moreover, the true impulsive limit $H_{free}{\ll}H_{pulse}$ is not currently accessible in the lab and therefore we cannot completely ignore free evolution while pulsing. As a result, while the above approach provides us with a framework and a starting point, we must employ quantum optimal control techniques, the details of which are achieved through numerical optimization, to apply the above methods for finite pulse lengths and finite coupling strength. We combined alternative steps of simulated annealing, which excels at extricating the search out of local minima but is slow at convergence to the optimum, with BFGS gradient following search, which excels at fast convergence but has an observed tendency of getting stuck at local minima.

Numerical studies for this work have been performed using QLib\cite{QLib}, a Matlab
package for quantum-information and quantum-optics calculations.
The pulses were optimized to give the lowest possible average number of phonons after cooling.

We optimized cooling cycles for a given initial temperature.
The optimization was performed in the impulsive limit
($\eta\Omega\gg\nu$), and applied for realistic parameter values,
where we may not neglect free evolution while pulsing. The sequences presented are just examples of what is achievable, and by no means are they to be considered
canonical in any way.

\bigskip{}

All computations below are done for the following physical settings,
which are achievable in the lab:
${}^{40}Ca^{+}$, $\nu=1MHz\cdot2\pi$, $\Omega=100MHz\cdot2\pi$,
$\lambda_{laser}=730nm$, giving a Lamb-Dicke value of $\eta=0.31$,
\cite{Monz}.
\begin{table}
\begin{tabular}{|l||c||c||c|}
\hline
 & Cycle A & Cycle B & Cycle C\tabularnewline
\hline
\hline
Initial energy [$\hbar\nu$] & 3 & 5 & 7\tabularnewline
\hline
Final energy [$\hbar\nu$] & 0.4 & 1.27 & 1.95\tabularnewline
\hline
Final energy after 25 cycles & 0.02 & 0.10 & 0.22\tabularnewline
\hline
Cycle duration & 4.4 $\frac{2\pi}{v}$ & 2.7 $\frac{2\pi}{v}$ & 0.8 $\frac{2\pi}{v}$\tabularnewline
\hline
No. pusles per cycle & 180 & 90 & 90\tabularnewline
\hline
No. of sequences & 10 & 10 & 10\tabularnewline
\hline
\end{tabular}\label{Flo:Cycles-data}

\caption{Cooling cycle performance. We define a cooling sequence as a series of alternating $\hat{X}\sigma_{x}$-pulses and $\hat{P}\sigma_{y}$-demi-pulses of varying lengths, followed by a reinitialization of the ion's internal dof. A cooling cycle is comprised of several cooling sequences, which are generally non-identical. }

\end{table}
Table 1 details 3 sample cycles and their cooling performance. Note
that the coupling terms in the propagator are proportional to $t_{p}\eta\sqrt{\frac{2m\nu}{\hbar}}\Omega\hat{X}\sigma_{\theta}$
and $\eta t_{f}t_{p}\Omega\nu\hat{P}\sigma_{y}$ and therefore total cooling time scales as $1/\sqrt{\Omega}$ since $t_p \times t_f$ scales like $1/\Omega$.

\begin{figure}
\includegraphics[scale=0.23]{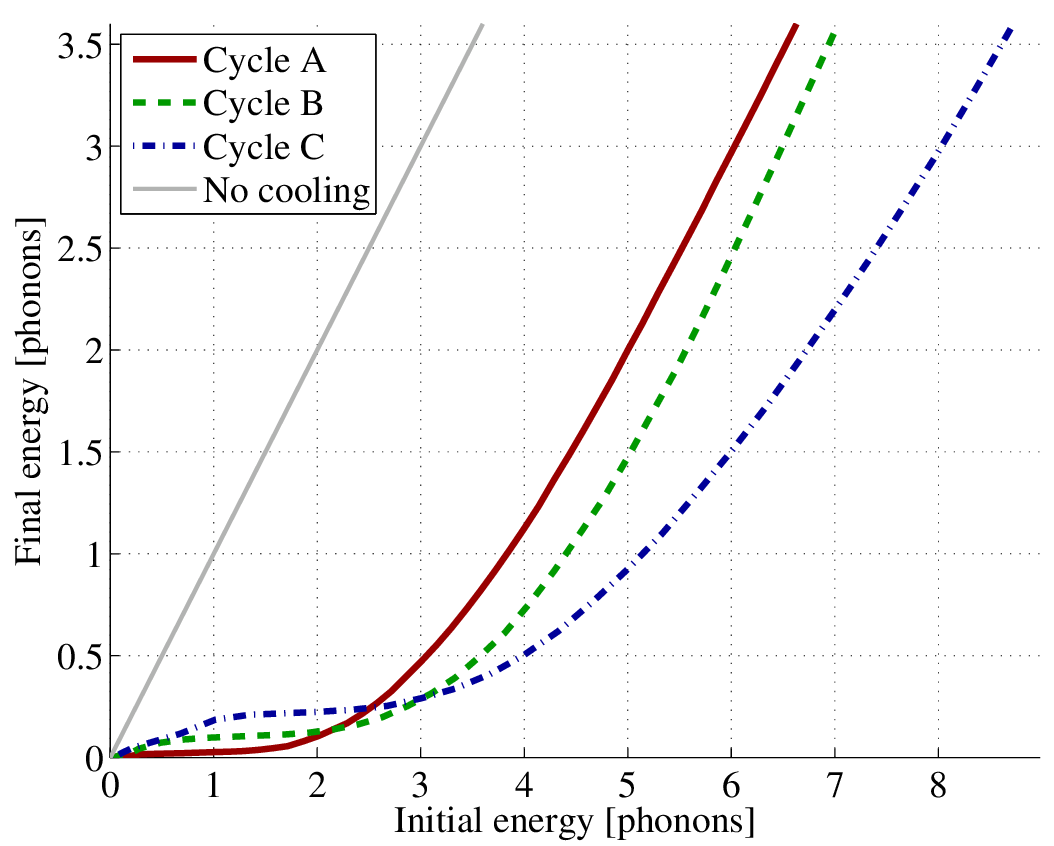}
\includegraphics[scale=0.23]{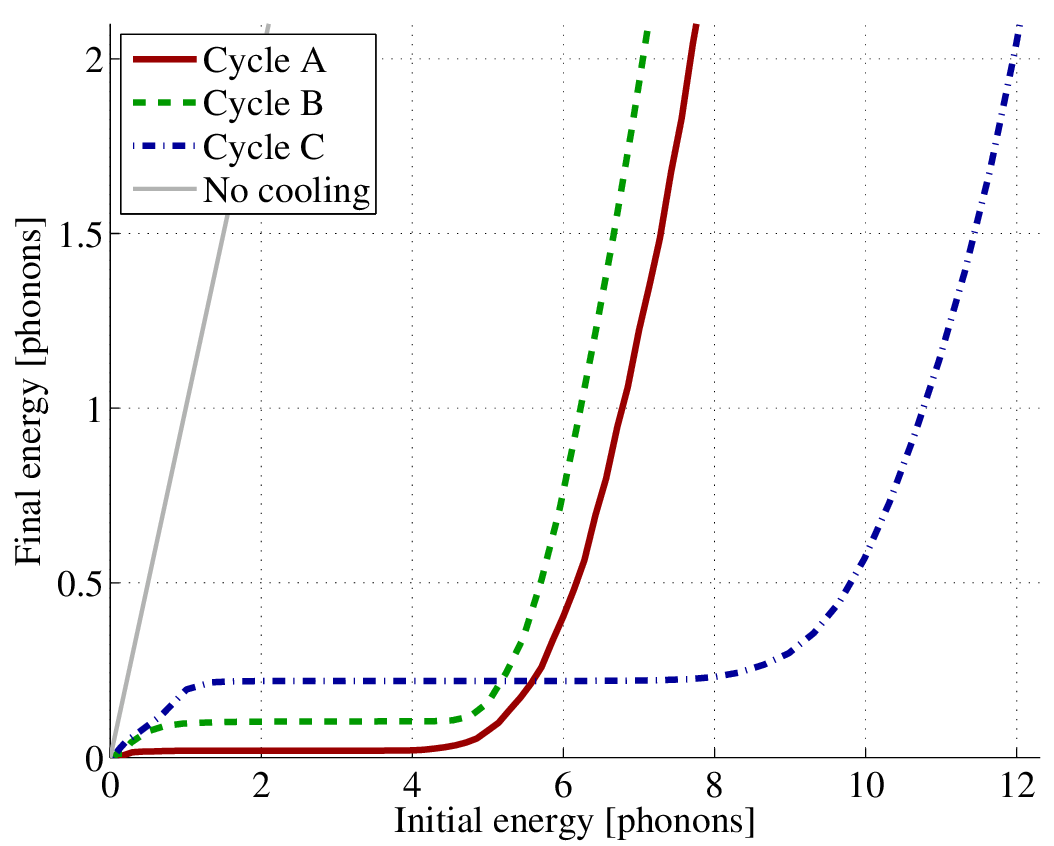}
\caption{Initial and final energy (above ground, in units of $\hbar\nu$) for a single application of the cooling
cycle and for 25 applications.}
\label{FinalInitial}
\end{figure}

We made the optimization in the impulsive limit, and then applied the deduced sequence using the full Hamiltonian. The results are shown in Fig. \ref{FinalInitial}. Starting from a thermal state of varying temperatures, Fig \ref{FinalInitial}a shows the final energy for a single application of a cooling cycle; Fig 1b shows the result of 25 applications of the same cycle. The latter cannot be deduced directly from Fig 1a because the state is no longer thermal after one or more cooling cycles. Note how the cycles give a good cooling for a wide range of initial temperatures, even though optimization was carried out for a specific initial temperature

\begin{figure}
\includegraphics[clip,scale=0.40]{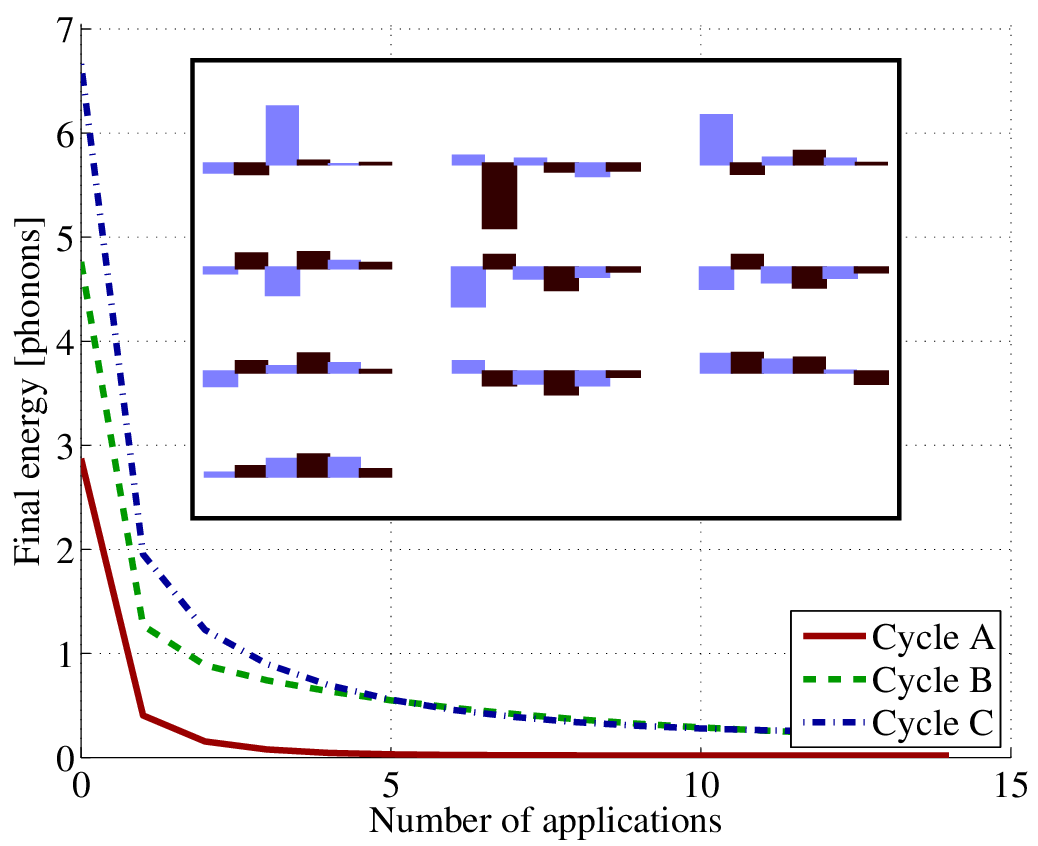}\label{Flo:Repeated applications}
\caption{Repeated applications of the sample cycles. {\\}Inset: X and P pulse lengths for sample cooling cycle C. Negative pulse lengths indicate the spin is to be oriented in the negative $x$ or $y$ directions. Each block of 6 bars (3 X and 3 P pulses) represent a single cooling sequence. These sequences are interspaced by reinitialization of the ion's internal state to make up a complete cooling cycle.}
\label{continuous}
\end{figure}

In Fig. \ref{continuous} we show how repeated applications of the three sample
cycles continues to lower the energy, until a steady state,
specific to each sequence is achieved.
An example timing of sequences within a cycle is shown in
fig. \ref{continuous} inset.

\begin{figure}
\includegraphics[clip,scale=0.4]{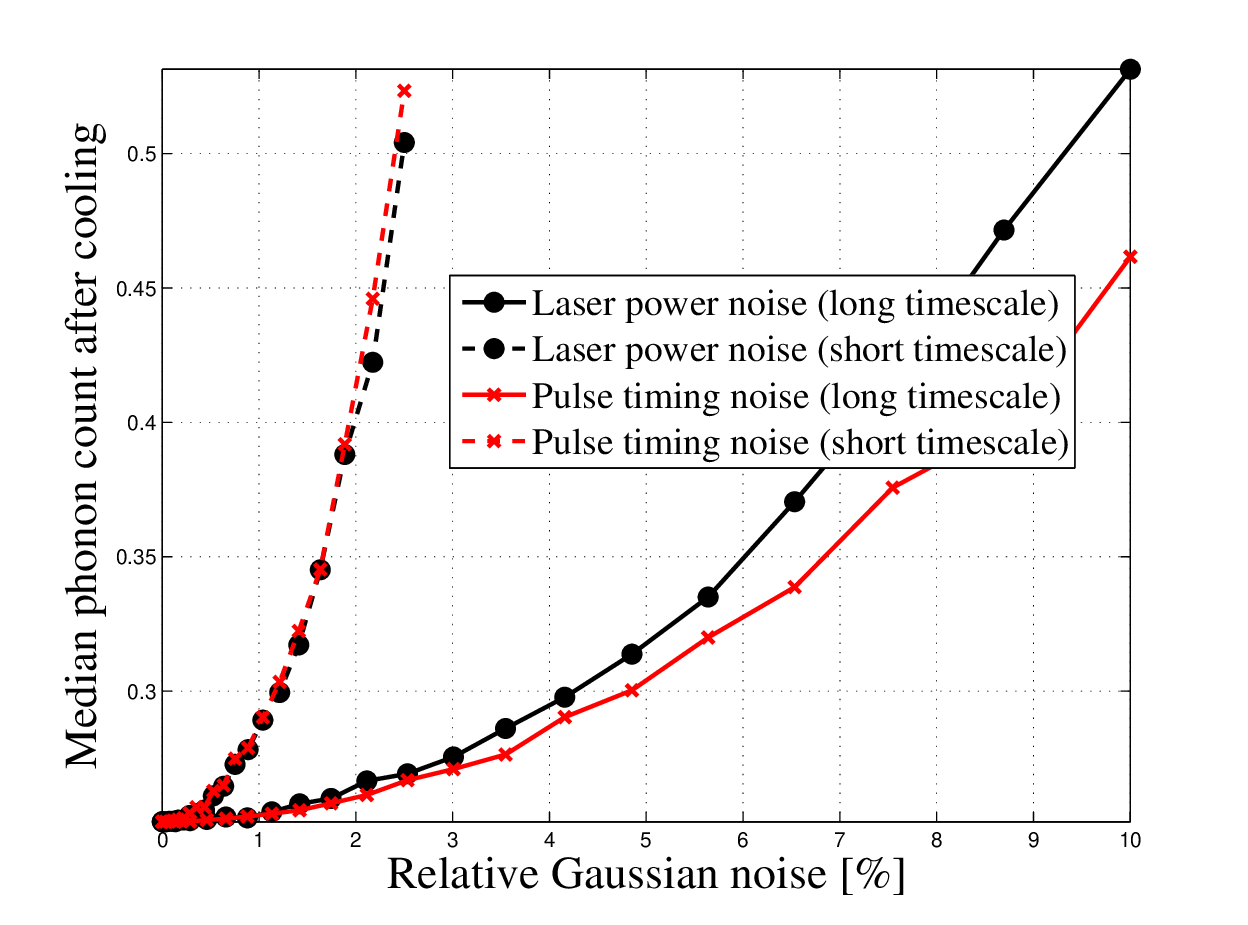}
\caption{Robustness to noise of cooling cycle C, applied 25 times. The x - axis represents the relative noise and the y- axis the final phonon population. As can be seen, superfast cooling is proven to be quite robust, both at long ($\nu$) and short ($\ll\nu$) timescales.}
\label{robust}
\end{figure}

{\em Robustness ---} The robustness of the cooling pulses is extremely important for experimental realization.
In order to analyze the robustness we have simulated the operation of the pulses under noisy conditions of the lasers. The results indicate that the cooling pulses are extremely robust even though they were not optimized to be so. In Fig. \ref{robust} we study the robustness of cycle C, repeated $25$ times,
to Gaussian noise in the pulse timings.
We assume that instead of the prescribed pulse time $t$, we implement
$t\longrightarrow\left(1+\epsilon\right)t$, with $\epsilon$ being
drawn from a Gaussian distribution with a varying standard deviation. A similar noise pattern was applied to the laser power, $\Omega$.
The superfast cooling exhibits two slightly different sensitivities to noise: for the short time-frame ($\ll\nu^{-1}$), which interferes with the commutation relations allowing the creation of the effective $P$ pulse, sensitivity is somewhat higher, while for noise occurring on longer timescales, i.e. the duration of the various pulses comprising the cooling sequence, Superfast cooling is extremely robust.
The plots show the mean final phonon count over $500$ cycles.
A graceful degradation in cooling performance can be observed.

Note that the state of the system after cooling is not a thermal state. This is due partly to purely numerical issues stemming from the need to exponentiate what are essentially infinite-sized Hamiltonians, and partly a physical feature of the specific cooling sequences presented. The latter results from the necessity to make use of relatively small density matrices in initial optimization stages, which resulted in effective cooling operators that do not cool high harmonic modes (above 30) as well as they cool lower modes. While this difference is negligible before cooling, when going significantly below $1$ phonon, the remaining energy in the high modes becomes a not-insignificant part of the overall energy. We believe that with higher computational power sequences resulting is more thermal-like final states are achievable.

 \bigskip{}

\begin{figure}
\includegraphics[clip,scale=0.6]{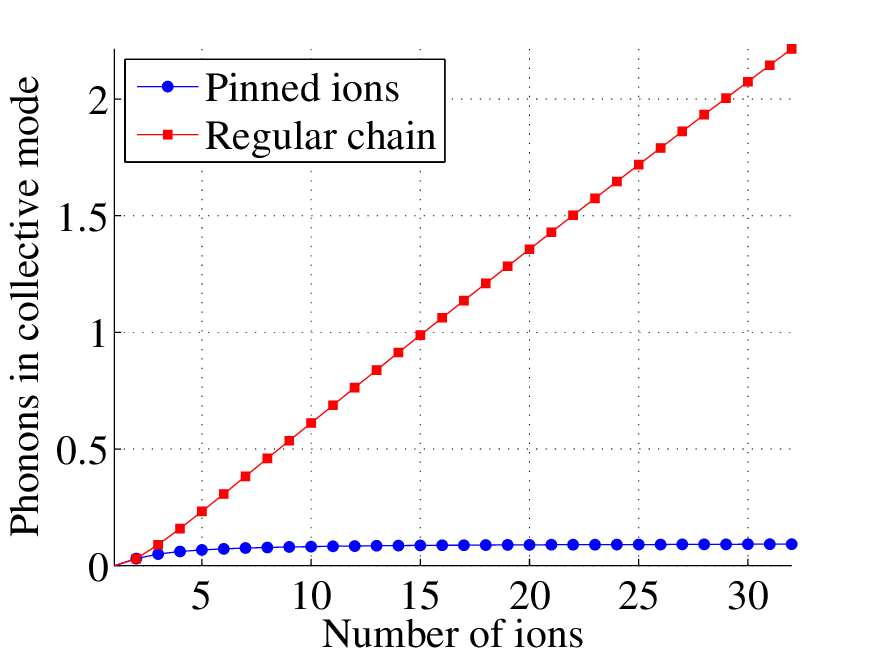}
\caption{The average population of the center of mass mode as a function of the number of ions, $\circ$ is for a regular chain and $\diamond$ is for pinned ions.
}
\label{chain}
\end{figure}

{\em Ion Chain ----}
By applying this cooling method to a large chain a product state of ground states can be reached, i.e., we will cool to the $\ket 0 \ket 0 .... \ket 0$ state in the local basis. This state is not far from the global ground state. In fig. \ref{chain} we show the number of phonons in the center of mass mode for a regular ion trap and for a trap in which all the ions are equidistantly pinned. In the latter case a saturation value of around 0.2 phonons is reached for the center of mass mode while for the regular trap there is no staturation but the population is still low for moderate chains.

{\em Standing wave ----}
There are a few ways to generate the standing wave Hamiltonian.
We are not really interested in a standing wave but in a pulse of a standing wave. We can emulate this pulse by running wave pulses by using the two following pulses:
$i\Omega \left( \sigma _{x}+\eta x\right),
-i\Omega \left( \sigma _{x}-\eta x\right)$.
These pulses can be created by passing a pulse through a beam splitter and
bringing it through two different directions, thus controlling the
phase and sign of the Lamb Dicke parameters independently.
The unitary transformation which describes this is:
$e^{i\Omega \sigma _{x}}e^{i\eta \Omega \sigma _{x}x}e^{-i\Omega \sigma
_{x}}e^{i\eta \Omega \sigma _{x}x}=e^{i2\eta \Omega \sigma _{x}x}$, which creates the unitary operation that we need. Moreover, since we are interested in fast cooling rates we also are ensured that the standing wave will not drift. The creation of the standing wave can be done as in\cite{Kilian}. In the case of mechanical oscillators this trick would not work since there is no temporal control of the effective Lamb Dicke parameter, but if the coupling is with Raman transitions a different intermediate level can be used and thus the effective Lamb Dicke will be different while the Rabi frequency may be the same and thus cancel the first pulse.

So far we have assumed the ion can be modeled by a non-dissipative
two-level system with effective Rabi frequency $\Omega$. This is
a simplification. Some experimental setups use a 3-level system with
the effective Rabi cycle achieved by way of a highly detuned Raman
transition off a highly-dissipative level \cite{Sideband-1,Sideband-2,Sideband-3}.
As a result, the true laser coupling of $100GHz$ is reduced to an
effective coherent $10MHz$ transition. However, unlike quantum information
processing, superfast cooling does not  require a fully
coherent transition. We could reduce the detuning and
achieve a higher Rabi frequency. At $\Omega=1GHz$, for example, the cooling process speeds up, requiring
only $0.4$, $0.3$ and $0.2\frac{2\pi}{\nu}$ for cycles A, B and
C, respectively.

\bigskip{}

{Nanomechanics --- }This scheme may have increasing importance in the nonmechanical and hydromechanical world since there the finite Q value limits the final temperature. One would use the setup described in \cite{Rabl2009,Ignacio2004}; i.e. couple a TLS to the oscillator. In this case the effective Lamb Dicke parameter is on the order of $\eta=0.03$(\cite{Ignacio2004}) and can be much higher for the scheme which is generated with magnetic gradients(\cite{Rabl2009}). In both of these schemes the strong coupling regime can be reached by strong lasers or even microwaves. The method currently in use  \cite{Florian2007,Ignacio2007}, assisted by a cavity, cannot beat the limit of the trap frequency due to the Gaussian nature of the scheme.

{\em Conclusion---}
To conclude, we have introduced the optimized-control approach to
cooling trapped ions, by optimizing over the realizable coupling
between the phonons and the ion's internal levels, $\hat{X}\sigma_{\theta}$,
and periods of free evolution. Surprisingly, even this meager palette is sufficient
to implement a red sideband-like operation and drive energy from the phonons
to the ions faster than the trapping frequency. Furthermore, the principles described
above are applicable to a very wide range of system and robust enough to be
implementable in ongoing experiments.

{\em Acknowledgments ---}
A.R. acknowledges the support of EPSRC
project number EP/E045049/1 and MBP acknowledges support
from the Royal Society and the EU STREP project HIP and the Alexander von Humboldt Professorship.


\begin{thebibliography}{11}
\bibitem{Hansch} T.W. Hänsch  and A.L. Schawlow, Optics Communications
{\bf 13}, 68,  1975.

\bibitem{Wineland1}  D. Wineland and H. Dehmelt, Bull. Am. Phys. Soc, {\bf 20}, 637 (1975).

\bibitem{SisyphusExp} P. Lett \etal \prl {\bf 61} 169 (1988).

\bibitem{SisyphusThe1} J. Dalibard \etal {\em Atomic Physics} {\bf 11}.

\bibitem{SisyphusThe2} J. Dalibard  and C. Cohen-Tannoudji {\em J. Opt. Soc. Am. B} {\bf 6}, 2023 (1989).

\bibitem{SisyphusBound} D. J. Wineland \etal  {\em J. Opt. Soc. Am. B} {\bf 9}, 32 (1992).

\bibitem{Aspect} A. Aspect \etal   {\em J. Opt. Soc. Am. B} {\bf 6}, 2112 (1989).

\bibitem{Kasevich} M. Kasevich and S. Chu \prl {\bf 69}, 1741 (1992).

\bibitem{MorigiDark} G. Morigi \etal \prl {\bf 85}, 4458 (2000).

\bibitem{RamanSideBandCooling} C. Monroe \etal, \prl {\bf 75}, 4011 (1995).

\bibitem{Rabl2009} P. Rabl \etal, \prb \textbf{79},  041302 (2009).

\bibitem{Ignacio2004} I. Wilson-Rae, P. Zoller, and A. Imamoglu, \prl \textbf{92},  075507 (2004).

\bibitem{QLib}S. Machnes, quant-ph/0708.0478, http://qlib.info

\bibitem{Monz} T. Monz \etal, \prl {\bf 103}, 200503 (2009).

\bibitem{Cold Ion Trap}D. Leibfried, R. Blatt, C. Monroe and D.J. Wineland.,  Rev. Mod. Phys. \textbf{75}, 281 (2003)

\bibitem{Lamb-Dicke regime}J. I. Cirac, L. M. Duan, and P. Zoller,  quant-ph/0405030

\bibitem{Suzuki-Trotter}M. Suzuki, Physica A \textbf{194}, 432 (1993)

\bibitem{Detecting in vacuum}A. Retzker, J. I.  Cirac, and B. Reznik, Phys. Rev. Lett. \textbf{94} 050504 (2005)

\bibitem{BCH-1}H. Baker, Proc Lond Math Soc (1) \textbf{34} 347\textendash{}360  (1902)
ibid (1) 35 (1903) 333\textendash{}374; ibid (Ser 2) \textbf{3} 24\textendash{}47 (1905)

\bibitem{BCH-2}J. Campbell, Proc Lond Math Soc \textbf{28} 381\textendash{}390 (1897)
ibid 29 (1898) 14\textendash{}32.

\bibitem{BCH-3} F. Hausdorff, Ber Verh Saechs Akad Wiss Leipzig \textbf{58} 19\textendash{}48 (1906)

\bibitem{BCH for 3}M.W. Reinsch, J. Math. Phys. \textbf{41}(4):2434
(2000).

\bibitem{Sideband-1}D.J. Wineland, R.E. Drullinger, F.L. Walls, Phys. Rev. Lett. \textbf{40} 1639 (1978)

\bibitem{Sideband-2}C. Monroe, \etal, Phys. Rev. Lett. \textbf{75},
4011 - 4014 (1995)

\bibitem{Sideband-3}V. Vuletic, C. Chin, A.J. Jerman, S. Chu, Phys. Rev. Lett. \textbf{81}, 5768 (1998)

\bibitem{Kilian} The mirror can be mounted to the electrodes of the ion trap as currnetly done in the Ion trap group in Ulm. Private communication with Kilian Singer.



\bibitem{Ignacio2007} I. Wilson-Rae \etal,  \prl {\bf 99},  093901 (2007).

\bibitem{Florian2007} F. Marquardt \etal, \prl {\bf 99},  093902 (2007).


\end{thebibliography}
\end{document}